\documentclass[journal=acsnano,manuscript=article,twocolumn]{achemso}
\usepackage[version=3]{mhchem} 
\usepackage{amsmath}
\usepackage{upgreek}
\usepackage{graphicx}
\usepackage[euler]{textgreek}
\usepackage[export]{adjustbox}
\usepackage{float}
\usepackage{physics}
\usepackage{mathrsfs}
\usepackage{bm}
\usepackage{caption}
\usepackage{textcomp}
\usepackage{natbib}
\usepackage{filecontents}
\usepackage{gensymb}
\usepackage{amsfonts} 
\usepackage{ragged2e}

\usepackage{adjustbox}
\usepackage{microtype} 
\usepackage{multicol}
\graphicspath{ {figures} }
\title{Impact of a MoS\textsubscript{2} monolayer on the nanoscale thermoelastic response of silicon heterostructures}

\begin{footnotesize}
\author{\footnotesize Davide Soranzio}
\affiliation{\footnotesize Institute for Quantum Electronics, Eidgenössische Technische Hochschule (ETH) Zürich, CH-8093 Zurich, Switzerland}
\email{davideso@phys.ethz.ch}
\author{Denny Puntel}
\affiliation{Dipartimento di Fisica, Università degli Studi di Trieste, Italy}
\author{Manuel Tuniz}
\affiliation{Dipartimento di Fisica, Università degli Studi di Trieste, Italy}
\author{Paulina E. Majchrzak}
\affiliation{Department of Physics and Astronomy, Interdisciplinary Nanoscience Center (iNANO), Aarhus University, 8000 Aarhus C, Denmark}
\author{Alessandra Milloch}
\affiliation{Department of Mathematics and Physics, Università Cattolica del Sacro Cuore, IT-25133 Brescia, Italy}
\alsoaffiliation{ILAMP (Interdisciplinary Laboratories for Advanced Materials Physics), Università Cattolica del Sacro Cuore, IT-25133 Brescia, Italy}
\alsoaffiliation{Department of Physics and Astronomy, KU Leuven, B-3001 Leuven, Belgium}
\author{Nicholas M. Olsen}
\affiliation{Department of Chemistry, Columbia University, New York, NY, USA}
\author{Wibke Bronsch}
\affiliation{Elettra - Sincrotrone Trieste S.C.p.A., Strada Statale 14, km 163.5, Trieste, Italy}
\author{Bjarke S. Jessen}
\affiliation{Department of Physics, Columbia University, New York, NY, USA}
\author{Danny Fainozzi}
\affiliation{Elettra - Sincrotrone Trieste S.C.p.A., Strada Statale 14, km 163.5, Trieste, Italy}
\author{Jacopo S. Pelli Cresi}
\affiliation{Elettra - Sincrotrone Trieste S.C.p.A., Strada Statale 14, km 163.5, Trieste, Italy}
\author{Dario De Angelis}
\affiliation{Elettra - Sincrotrone Trieste S.C.p.A., Strada Statale 14, km 163.5, Trieste, Italy}
\author{Laura Foglia}
\affiliation{Elettra - Sincrotrone Trieste S.C.p.A., Strada Statale 14, km 163.5, Trieste, Italy}
\author{Riccardo Mincigrucci}
\affiliation{Elettra - Sincrotrone Trieste S.C.p.A., Strada Statale 14, km 163.5, Trieste, Italy}
\author{Xiaoyang Zhu}
\affiliation{Department of Chemistry, Columbia University, New York, NY, USA}
\author{Cory R. Dean}
\affiliation{Department of Physics, Columbia University, New York, NY, USA}
\author{Søren Ulstrup}
\affiliation{Department of Physics and Astronomy, Interdisciplinary Nanoscience Center (iNANO), Aarhus University, 8000 Aarhus C, Denmark}
\author{Francesco Banfi}
\affiliation{Université de Lyon, CNRS, Université Claude Bernard Lyon 1, Institut Lumière Matière, F-69622 Villeurbanne, France}
\author{Claudio Giannetti}
\affiliation{Department of Mathematics and Physics, Università Cattolica del Sacro Cuore, IT-25133 Brescia, Italy}
\alsoaffiliation{ILAMP (Interdisciplinary Laboratories for Advanced Materials Physics), Università Cattolica del Sacro Cuore, IT-25133 Brescia, Italy}
\alsoaffiliation{CNR-INO (National Institute of Optics), via Branze 45, IT-25123 Brescia, Italy}
\author{Fulvio Parmigiani}
\affiliation{Elettra - Sincrotrone Trieste S.C.p.A., Strada Statale 14, km 163.5, Trieste, Italy}
\alsoaffiliation{Dipartimento di Fisica, Università degli Studi di Trieste, Italy}
\alsoaffiliation{International Faculty, University of Cologne, Albertus-Magnus-Platz, 50923 Cologne, Germany}
\author{Filippo Bencivenga}
\affiliation{Elettra - Sincrotrone Trieste S.C.p.A., Strada Statale 14, km 163.5, Trieste, Italy}
\author{Federico Cilento}
\affiliation{Elettra - Sincrotrone Trieste S.C.p.A., Strada Statale 14, km 163.5, Trieste, Italy}
\email{federico.cilento@elettra.eu}

\end{footnotesize}

\begin{document}
\singlespacing
\clearpage
\twocolumn[
  \begin{@twocolumnfalse} 
    \maketitle 
    \begin{center}
        \Large{Abstract}
    \end{center}
Understanding the thermoelastic response of a nanostructure is crucial for the choice of materials and interfaces in electronic devices with improved and tailored transport properties, at the length scales of the present technology. Here we show how the deposition of a MoS\textsubscript{2} monolayer can strongly modify the nanoscale thermoelastic dynamics of silicon substrates close to their interface.
We achieve this result by creating a transient grating with extreme ultraviolet light, using ultrashort free-electron laser pulses, whose $\approx$84 nm period is comparable to the size of elements typically used in nanodevices, such as electric contacts and nanowires.
The thermoelastic response, featured by coherent acoustic waves and an incoherent relaxation, is tangibly modified by the presence of monolayer MoS\textsubscript{2}. Namely, we observed a major reduction of the amplitude of the surface mode, which is almost suppressed, while the longitudinal mode is basically unperturbed, aside from a faster decay of the acoustic modulations. We interpret this behavior as a selective modification of the surface elasticity and we discuss the conditions to observe such effect, which might be of immediate relevance for the design of Si-based nanoscale devices. 
\vspace{1 cm}    
  \end{@twocolumnfalse}
]

\noindent
Transition metal dichalcogenides (TMDs) are a class of materials composed of atomic layers held together by van der Waals interactions, much weaker than the intralayer covalent bonds, giving them a marked two-dimen- sional character. This feature allows to obtain controlled thicknesses down to the single layers, useful for tuning several physical properties, such as the electronic band gap, the vibrational levels and the excitons, and for their implementation in thin nanodevices, \textit{e.g.}, transistors, photodetectors and electroluminescent devices  \cite{Lee2010,Wang2012,Golovynskyi2020,Ellis2011}{}. In this respect, a key point is the interaction between TMDs and substrates, which has been shown to critically impact the heterostructure properties. Hence, the choice of the substrate constitutes a fundamental aspect for the application of TMDs heterostructures in electronic circuits \cite{Gabourie2022,Li2020,Agmon2022,Kuppadakkath2022}{}. 
To design functional devices, the thermoelastic response of the structure, \textit{i.e.}, how fast can heat be dissipated, together with its elastic properties, which are also relevant for technological applications \cite{Hartmann1985,Biryukov1995}{}, are fundamental aspects to be considered. 

In this work, we focus on the dynamics over the typical length-scales of components used in nanodevices, \textit{e.g.}, electric contacts and nanowires \cite{Bergin2012,Leong2015}{}.
In particular, we study the nanoscale thermoelastic response of a TMD monolayer (ML) deposited on top of a silicon substrate. The selected TMD material is molybdenum disulfide (MoS\textsubscript{2}), which presents an interlayer separation of 0.65 nm and a $\approx$1.9 eV direct band gap at the ML limit \cite{Ganatra2014}{}.
To investigate the thermoelastic dynamics, we used the transient grating (TG) technique \cite{Nelson1982}{}. Specifically, nanoscale TGs were generated on the sample using free electron laser (FEL) extreme-ultraviolet (EUV) femtosecond pulses, to attain a $\approx$84 nm grating period. 
We studied the TG dynamics in three systems: a blank Si membrane, a ML MoS\textsubscript{2} transferred on a Si membrane and a ML MoS\textsubscript{2} grown on a Si wafer.
The results showed a marked difference in the surface thermoelastic response of Si when ML MoS\textsubscript{2} is added. While two acoustic modes, a superficial Rayleigh-like and a longitudinal, were detected for the blank membrane, the longitudinal mode dominates for the MoS\textsubscript{2}-covered samples alongside with a faster decay of the modulations.
We ascribe this fact to the different atomic motion involved, which is elliptical and localized close to the interface for the surface mode and parallel to the interface for the longitudinal mode.
Underneath the acoustic modulations, we observed a slower decay of the non-oscillating background, usually assigned to thermal transport \cite{Bencivenga2019}{}, for the heterostructures compared to the blank substrate.

\section{Results and discussion}

\begin{figure*}[t]
\includegraphics[width=\textwidth]{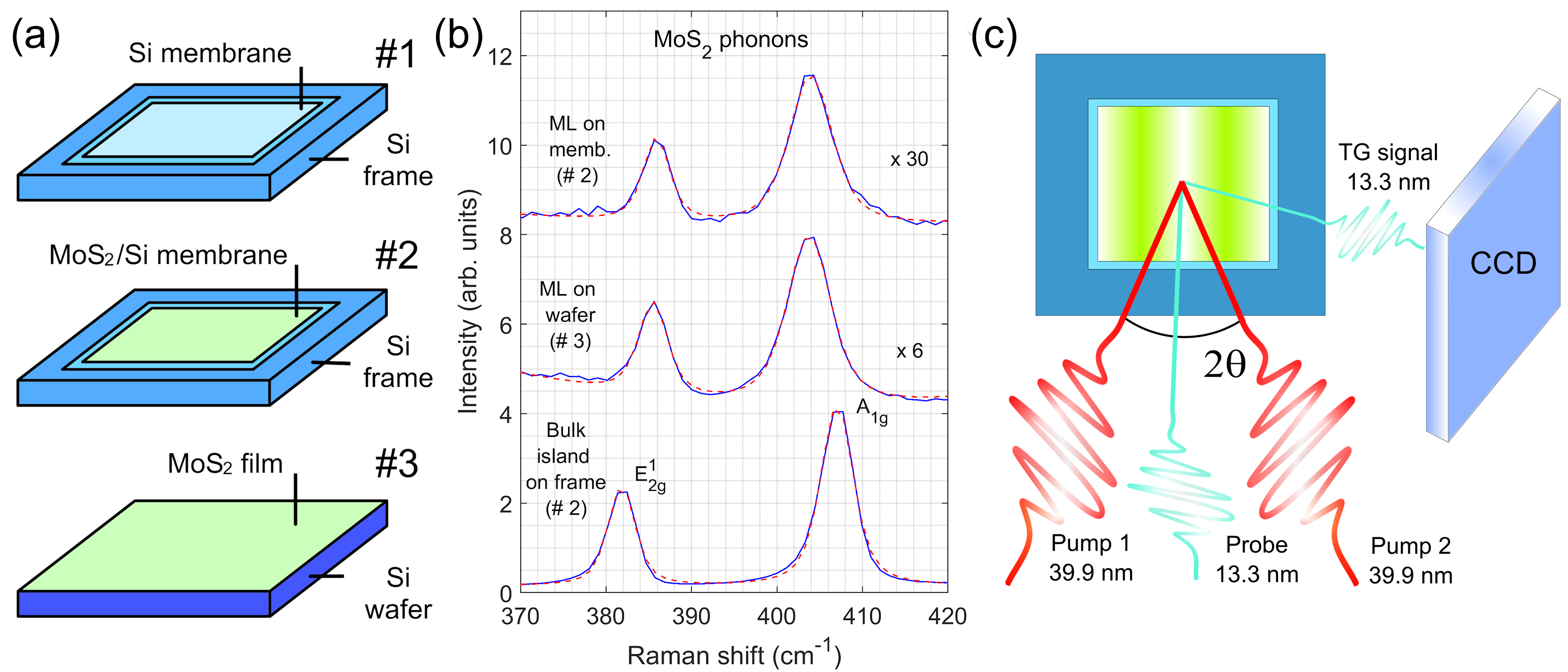}

\caption{(a) Samples studied during the TG experiments: \#1 blank Si membrane, \#2 MoS\textsubscript{2}/Si membrane and \#3 MoS\textsubscript{2}/Si wafer. (b) Selected micro-Raman spectra showing  the $E^1_{2g}$ and $A_{1g}$ phonon peaks from the investigated samples: MoS\textsubscript{2}/Si membrane, MoS\textsubscript{2}/Si wafer and a MoS\textsubscript{2} bulk island. The Raman traces were rescaled and vertically shifted for clarity. The blue lines correspond to the experimental data, while the red-dotted ones to the best fit. (c) Scheme of the TG setup at the EIS-TIMER beamline at FERMI. }

 \label{fig1}
\end{figure*}

A schematic of the samples is shown in Fig. \ref{fig1}(a). A blank Si membrane was used as a reference (sample \#1). On top of a nominally identical membrane, a large-area monolayer of MoS\textsubscript{2} was prepared using the gold tape method (sample \#2) \cite{Liu2020}{}.
For comparison to the membrane systems, we also studied a CVD-grown ML MoS\textsubscript{2}, deposited on a Si wafer (sample \#3). More details regarding the sample preparation are reported in the Methods section.

The MoS\textsubscript{2} coverage was tested using micro-Raman, taking advantage of the relation between the frequencies of the MoS\textsubscript{2} $E^1_{2g}$ and $A_{1g}$ near-zone-center phonon modes and number of layers \cite{Lee2010}{}. Sample scans are reported in Fig. \ref{fig1}(b), with the ML-covered regions from sample \#2 and \#3 showing a much smaller ($\approx$19 cm\textsuperscript{-1}) frequency separation of the modes, when compared to a thick multilayer, bulk-like island found on the Si frame ($\approx$25 cm\textsuperscript{-1}), which is consistent with the literature.
A systematic characterization of the samples is reported in the Supplemental Material \cite{supplementary}{}. Using the bulk MoS\textsubscript{2} film island as a reference, we conclude that the values found for both sample \#2 and \#3 are compatible with a ML coverage.

\begin{figure*}[h!]
\includegraphics[width=\textwidth]{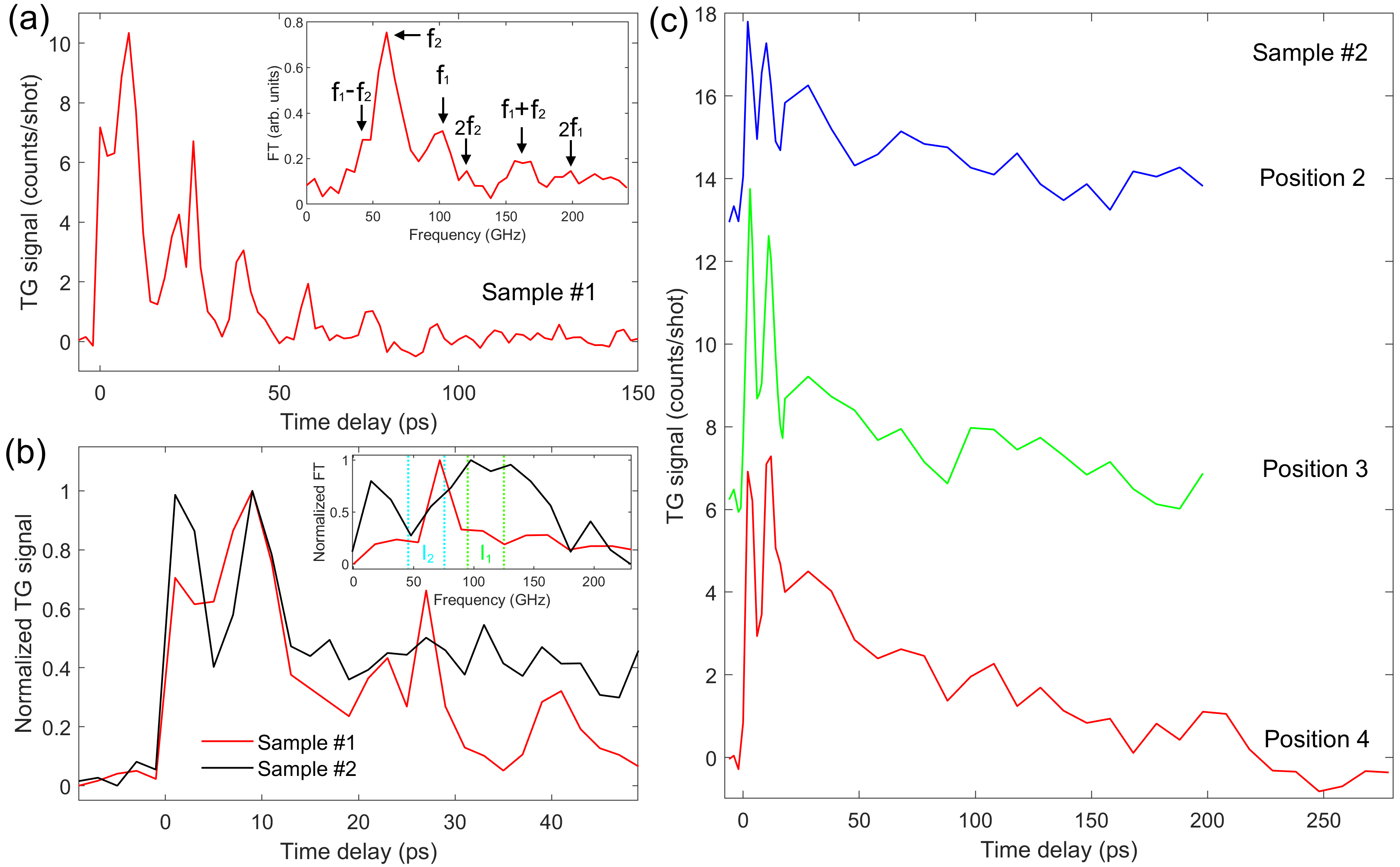}

\caption{(a) TG signal from the blank Si membrane (sample \#1) at 0.34 mJ/cm\textsuperscript{2} fluence; the inset shows the Fourier transform of the signal after the subtraction of an exponentially-decaying background for the $\approx$160 ps window. (b) TG signal from the MoS\textsubscript{2}/Si membrane (sample \#2) at 0.23 mJ/cm\textsuperscript{2} fluence compared to the one from (a); the inset shows the Fourier transform of the two signals after the subtraction of an exponentially-decaying background for the $\approx$60 ps window; two integration region are marked (see text)
(c) TG signal collected at different positions compared to (b) on the membrane of sample \#2 under 0.25 mJ/cm\textsuperscript{2} fluence.}

 \label{fig2}
\end{figure*}

Fig. \ref{fig1}(c) illustrates the EUV TG experimental geometry. Two symmetric pump beams with wavelength $\lambda_{pu}=39.9$ nm and time duration $<$ 100 fs cross at the sample with an angle $2\theta$=27.6$^{\circ}$ leading to a TG with a periodicity $\Pi=2\pi/k_{TG}$=83.6 nm, where $k_{TG}=4\pi sin(\theta)/\lambda_{pu}$ is the modulus of the TG vector. The probe beam was set at $\lambda_{pr}=13.3$ nm. More details regarding the EUV TG setup are provided in the Methods section.
All the measurements were performed in reflection geometry, in a high vacuum and at room temperature. The diffracted signal was recorded using a CCD camera.

Fig. \ref{fig2}(a) shows the time-resolved TG signal for the blank Si membrane (sample \#1). Before the temporal overlap between pump and probe, occurring at $t$=0, no signal is detected, while for $t>0$ a transient diffracted signal appears. It is featured by modulations on top of the decay of the average signal, which is typical for a thermoelastic response.

To analyze the TG signal, we model its time-resolved intensity using the expression \cite{Bencivenga2019}{}
\begin{multline}
 I(t)=\theta(t)\cdot\\
 \left |A_{th}e^{-t/\tau_{th}}+\sum^n_{i=1}A_{i}e^{-t/\tau_{i}}\cos\left(2\pi f_{i}t+\phi_i\right)\right|^2
 \label{eq2}
 \end{multline}

\noindent
where $\theta(t)$ is the Heaviside function, $A_{th}$ and $\tau_{th}$ are the amplitude and decay time usually assigned to the thermal component, $A_{i}$, $f_{i}$, $\tau_{i}$ and $\phi_{i}$ are the amplitude, frequency, time decay constant and phase of the \textit{i}-th mode.
\noindent
Two modes are sufficient to describe the TG signal from  sample \#1. In fact, the square in the formula leads, when expanded, to a series of exponentially-decaying terms oscillating at frequencies 0, $f_1$, $f_2$, $2f_1$, $2f_2$, $f_1+f_2$ and $f_1-f_2$.
The non-null frequency contributions can be visualized by fitting the experimental data to Eq. \ref{eq2} with the oscillatory amplitudes $A_i$ set to zero and performing the discrete Fourier transform on the residual signal (inset of Fig. \ref{fig2}(a)). 
The result is consistent with two modes with overtones, corresponding to the sums and differences of two frequencies $f_1$ and $f_2$, such that $f_1\approx 102$ GHz, $f_2\approx 60$ GHz, $2f_1\approx 204$ GHz, $2f_2\approx 120$ GHz, $f_1+f_2\approx162$ GHz and $f_1-f_2\approx42$ GHz, with the last one appearing as the left shoulder of the most prominent peak at $f_2\approx60$ GHz. 
By fitting the data using the full Eq. \ref{eq2}, we obtain the exact frequencies of the two modes, $f_1=(105\pm1)$ GHz and $f_2=(58.2\pm0.4)$ GHz. Moreover, the decay time is found to be $\tau_{th}=(39\pm2)$ ps.

\begin{figure}[h!]
\includegraphics[width=\columnwidth]{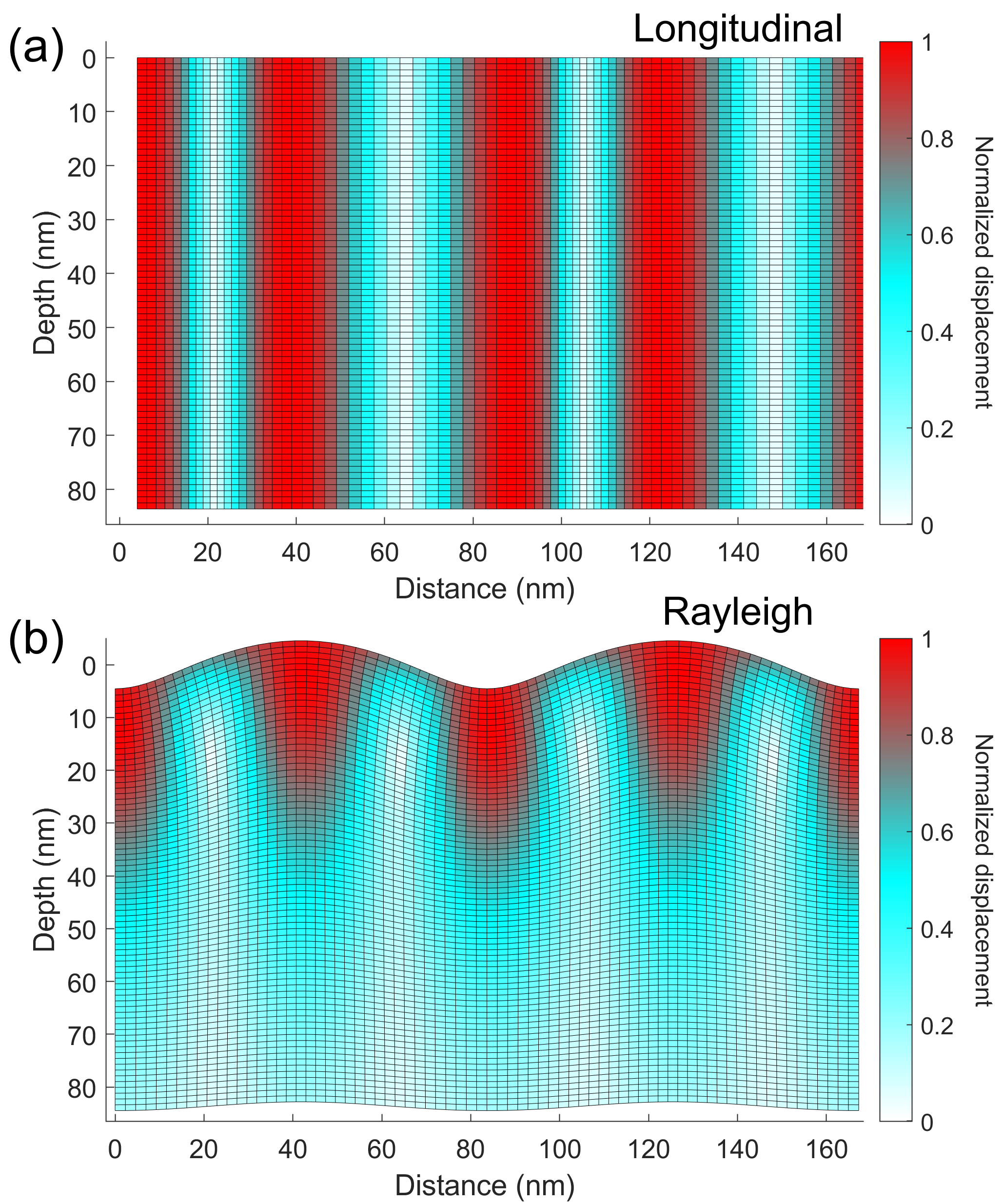}

\caption{Schematic representation of the displacements for the (a) longitudinal (b) Rayleigh acoustic waves in silicon. Their amplitude was arbitrarily set by a common factor to ease visualization; in the colorbar, the displacements were normalized to the maximum one with respect to the equilibrium positions.}

 \label{fig23}
\end{figure}

\begin{figure*}[h!]
\includegraphics[width=\textwidth]{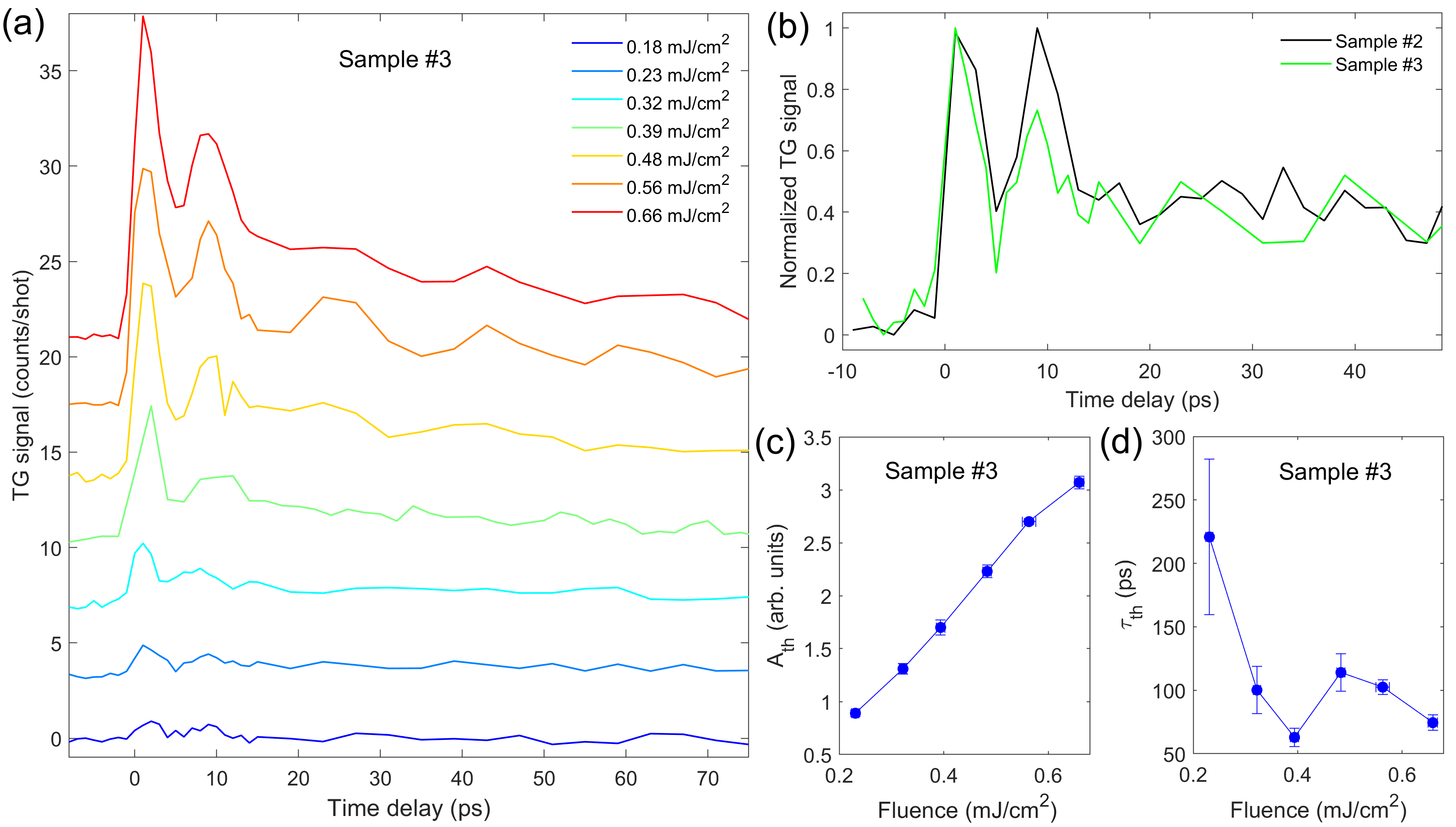}

\caption{TG signal from sample \#3  (a) Fluence dependence of TG signal from sample \#3. (b) Comparison between the normalized TG signals from sample \#2 (0.23 mJ/cm\textsuperscript{2}) and sample \#3 (0.25 mJ/cm\textsuperscript{2}). (c) Amplitude and (d) time decay constant of the non-oscillating component, as derived by the best fit of EUV TG waveforms from panel (a) to Eq. \ref{eq2}.}

 \label{fig3}
\end{figure*}

We can rationalize the presence of these two modes as Lamb waves, namely, acoustic waves of the membrane \cite{Auld1973_2,Royer2022}{}. In TG experiments, their momentum is set by $k_{TG}$, while the propagation speed is related to the frequency-momentum dispersion relation of the material. Using the frequencies derived from the TG signal, it is possible to obtain the corresponding phase velocities with $v_{p,i}=f_i\Pi$. For sample \#1, the detected modes correspond to $v_{p,1}=(8.81\pm0.09)$ km/s and $v_{p,2}=(4.87\pm0.03)$ km/s. These can be compared to the predicted phase velocity dispersion for the 200-\textmu m-thick Si membrane, through the transverse resonance method \cite{Auld1973_2,mastersthesis}{}, and using the longitudinal and shear velocities reported in Farnell and Adler \cite{Farnell1972}{}. The resulting phase velocities depend on the product $\rho=k_{TG}b$, where, in the present case, $k_{TG}$=0.075 nm\textsuperscript{-1} and $b=200$ nm is the membrane thickness; hence $\rho\approx15$. At this $\rho$ value, two modes with phase velocities 8.89 km/s and 4.90 km/s, close to the longitudinal and transverse ones for Si, are predicted, compatible with the observed ones. The fitting procedure also provides the acoustic decay constants, which result $\tau_1=(80\pm50)$ ps and $\tau_2\approx470$ ps. Both values are either comparable or larger than the investigated delay range, as shown in Fig. \ref{fig2}(a), hence a more accurate estimation would require further dedicated measurements.

Similar results have been previously reported for optical and EUV TG experiments in other materials for $\rho\gg1$ \cite{Janušonis2016,Maznev2018,Brioschi2023,Foglia2023,Bencivenga2023}{}. The acoustic modes are commonly identified as the `surface-skimming' longitudinal wave (SSLW) and, in the limit of a semi-infinite substrate ($\rho\to\infty$), the Rayleigh surface acoustic wave (RSAW). As observed in our measurements, the SSLW at $f_1$=105 GHz is only visible in the first $\approx$15 ps in the TG signal, since it behaves like a leaky wave which rapidly decays in the bulk \cite{Janušonis2016}{}. The RSAW decays on a longer timescale and involves elliptical displacements, \textit{i.e.}, combined longitudinal and transversal motion. These are approximately localized at the surface within half a  wavelength \cite{Janušonis2016_prb,Royer2022}{}. A depiction of the two waves is given in Fig. \ref{fig23}, based on the derivation of their displacements with the method of potentials \cite{Royer2022}{}; more details are provided in the Supplementary Information \cite{supplementary}{}.
The accurate prediction of the wave and heat propagation is complex and depends on many experimental details, such as the energy distribution in time and space. Nevertheless, in the reflection configuration used in the experiments, one expects the dominant contribution to the TG signal to arise from a coherent surface displacement, i.e., from the RSAW \cite{Xu2004,Veres2013,Janušonis2016,Liu2021}{}.

In Fig. \ref{fig2}(b) we show the time-resolved TG signal for the MoS\textsubscript{2}/Si membrane heterostructure (sample \#2) compared to the blank membrane (sample \#1). Differently from the blank case, here we observe an oscillatory signal in the first $\approx$15 ps only, with two prominent peaks. By performing the discrete Fourier transform after subtracting the non-oscillatory signal from the two traces as for panel (a), we observe a clear difference between sample \#1 and sample \#2 as shown in the inset in panel (b). In fact, if we integrate the spectral amplitude around the two main mode frequencies ($f_1$, 90-120 GHz) and ($f_2$, 45-75 GHz), we obtain that their ratio is I\textsubscript{1}/I\textsubscript{2}$\approx$0.6 for sample \#1 and $\approx$1.9 for sample \#2, pointing to a marked difference in the relative contribution of the two modes.
Afterwards, the signal gradually decays with minor modulations on a longer timescale with respect to the blank substrate (Fig. \ref{fig2}(a),(c)). Repeated scans at different sample positions gave an analogous response.

Fig. \ref{fig3}(a) shows a fluence dependence of the TG signal collected on the MoS\textsubscript{2}/Si wafer (sample \#3).
Analogously to sample \#2, we observe a response mainly localized in the first picoseconds with two main oscillation periods, resembling the SSLW detected in sample \#1, followed by lower frequency modulations that emerge only for the highest investigated fluences, similar to the ones associated to the RSAW observed for sample \#1.
The TG signal acquired at different positions on the sample showed analogous features, albeit with some amplitude variability possibly connected to the local MoS\textsubscript{2} coverage (see Supplementary Information \cite{supplementary}{}).
A comparison between the response of samples \#2 and \#3 under similar excitation conditions is reported in Fig. \ref{fig3}(b), showing similar features in the dynamics.

Considering the values of the attenuation length for Si and MoS\textsubscript{2}, reported in the Methods section (at 39.9 nm, $\approx$237 nm and $\approx$40 nm respectively), it is evident that a MoS\textsubscript{2} ML cannot significantly alter the amount of energy deposited in the Si substrates. On the other hand, the monolayer/substrate interaction leads to a thermal resistance and changes in the dynamical heat distribution and elasticity at the interface of MoS\textsubscript{2}/Si heterostructures; these properties are also influenced by the surface nanogroove of the substrate which is connected dynamically to the thermoelastic response \cite{Wang2021,Liu2023}{}. 
Therefore, one may expect surface modes like the RSAW, involving both longitudinal and transverse displacements \cite{Royer2022}{}, to be more affected by the interaction with the MoS\textsubscript{2} ML rather than the longitudinal wave that leaks into the bulk. This reasoning is consistent with the experimental results reported in Figs. \ref{fig2}(b),(c) and \ref{fig3}(a), where a main oscillatory contribution with frequency close to the longitudinal mode dominates the TG signal of the samples \#2 and \#3, where a ML MoS\textsubscript{2} is present.

To give a more quantitative view, we fitted the data from sample \#2 and \#3 using Eq. \ref{eq2}. Besides the TG signal collected from sample \#3 at the highest fluences, the resolved oscillations are limited in time to about two periods of the longitudinal mode. This does not give enough data to effectively disentangle the contributions at distinct frequencies due to the correlation among the parameters, especially for the lower fluences. Nonetheless, it is possible to give good estimates keeping some of the parameters fixed. 
For sample \#2, fixing the frequencies and phases of the two modes from the fit to the sample \#1 TG signal (Fig. \ref{fig2}(a)), we obtained an average non-oscillating background decay constant $\tau_{th}=(170\pm10)$ ps over the different positions on the sample at 0.25 mJ/cm\textsuperscript{2} (Fig. \ref{fig2}(c)). 
For sample \#3, we first model the response at 0.56 mJ/cm\textsuperscript{2} (Fig. \ref{fig3}(a)), where both modes emerge, keeping all the parameters free. The resulting frequencies are $f_1=(113\pm5)$ GHz and $f_2=(55\pm1)$ GHz, which are connected to the velocities $v_1=(9.4\pm0.4)$ km/s and $v_2=(4.62\pm0.09)$ km/s, close to the ones of the membrane. Regarding the decay constant of the acoustic modes, we obtained $\tau_1=(5\pm1)$ ps and $\tau_2=(70\pm40)$ ps. 
We then fix the frequencies, phases and time decay constants of the acoustic modes to extract the fluence dependence of the amplitude $A_{th}$ and time decay constant $\tau_{th}$, which we report in Fig. \ref{fig3}(c),(d). We omit the fit results for the lowest fluence as the time decay parameter is heavily influenced by the lower signal-to-noise ratio of this trace. While $A_{th}$ increases linearly in the explored fluence range, $\tau_{th}$ settles around $\approx$100 ps for most of the explored range. The higher value $\tau_{th}=(220\pm 60)$ ps from sample \#3 at 0.23 mJ/cm\textsuperscript{2} may appear as an outlier, however it is compatible with the value derived from the sample \#2 data under close excitation conditions ($\tau_{th}=(170\pm10)$ ps at 0.25 mJ/cm\textsuperscript{2}).

\begin{figure}[h!]
\includegraphics[width= 6.9 cm]{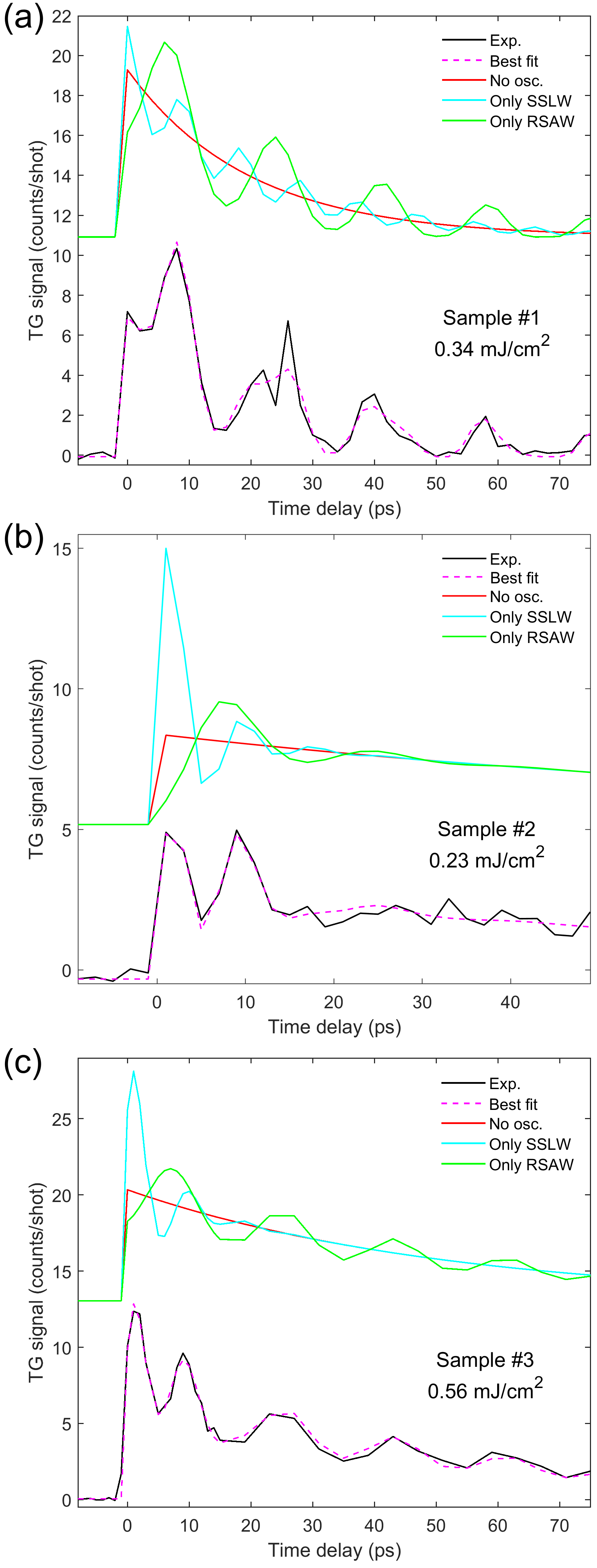}

\caption{Contribution of the acoustic modes to the TG signal illustrated by comparing the experimental data (Exp.) to the best-fit curve using Eq. \ref{eq2} (Best fit) and by selectively removing their contribution from the fit result (only SSLW, only RSAW and no osc.) for samples (a) \#1 (b) \#2 (c) \#3.}

 \label{fig5}
\end{figure}

To illustrate how the effects of the two acoustic modes combine differently between the blank Si membrane and the MoS\textsubscript{2} heterostructures, in Fig. \ref{fig5} we present examples from the three samples: (a) \#1 at 0.34 mJ/cm\textsuperscript{2}, (b) \#2 at 0.23 mJ/cm\textsuperscript{2} and (c) \#3 at 0.56 mJ/cm\textsuperscript{2}.
Here we remove from the fit (indicated with `Best fit'), obtained using  Eq. \ref{eq2}, the contribution of one (`Only SSLW' or `Only RSAW') or both (`No osc.') of the two acoustic modes, by setting their amplitude to zero. It is clear that the double-peak structure found in the MoS\textsubscript{2} heterostructures originates from the superposition between longitudinal and Rayleigh oscillations, where the relative height of the first two peaks is dictated by the combination of amplitude and phase of the two modes, more rapidly damped than the sample \#1 case.

The observation of the influence from the ML MoS\textsubscript{2} on the TG signal is favoured by the choice of the substrate. We see from Fig. \ref{fig3}(a) that the amplitudes of the acoustic modes increase with the pump fluence, that determines larger displacements from the atomic equilibrium positions. The attenuation length of our pump beams is $\approx$237 nm, comparable to the membrane thickness $b\approx200$ nm for sample \#2. In another commonly employed substrate for TMDs such as sapphire, the attenuation length is just $\approx$13 nm, about 18 times smaller \cite{supplementary}{}. This gives a much higher excitation energy density than in the case of Si, potentially giving much larger displacements due to the RSAW mode, which are localized within a depth of half of the mode wavelength in the material (Fig. \ref{fig23}(b)). For fluences comparable to the ones employed for the data reported in Figs. \ref{fig2} and \ref{fig3}, the RSAW in the TG signal dominates the response also in a ML WS\textsubscript{2}/Sapphire heterostructure. We provide an example of this in the Supplemental Material \cite{supplementary}{}.

Finally, we discuss the decay of the non-oscillatory component of the TG signal. On sample \#1, we  obtained $\tau_{th}=(39\pm2)$ ps. This value is substantially slower than what is expected by the standard thermal diffusive theory. Using the Si thermal diffusivity $\alpha_S\approx 0.055$ \textmu{}m\textsuperscript{2}/nS from Johnson et al. \cite{Johnson2013}{} and our grating vector $k_{TG}= 0.075$ nm\textsuperscript{-1}, the thermal diffusion theory predicts a decay $\tau_{Si}=1/\alpha_s k^2_{TG}\approx$3.2 ps, about one order of magnitude smaller than the value from our experiment. This deviation was previously described as a consequence of ballistic transport, occurring for $k_{TG}\Lambda_{mfp}\gg1$, where $k_{TG}$ is the TG vector and $\Lambda_{mfp}$=0.5-1 \textmu m is the median phonon mean free path \cite{Johnson2013,Maznev2011}{} .

With the addition of ML MoS\textsubscript{2}, we observed a marked increase in the non-oscillatory time decay in samples \#2 and \#3. 
However, the decay constant found for the heterostructure is almost as large or higher than the investigated temporal windows for sample \#2 and \#3 (Fig. \ref{fig2}(b),(c) and \ref{fig3}(a)). Therefore, a further dedicated set of measurements is required for an accurate explanation of this result. Based on the considerations for Si, it might be important to take into consideration a model that includes non-diffusive thermal transport, the interfacial thermal resistance in the MoS\textsubscript{2} and Si layers \cite{Gabourie2022,Liu2023}{} and possibly the influence of charge recombination at the MoS\textsubscript{2}/Si interface \cite{Zhao2019}{}. Furthermore, as we have shown in our work, the amplitude and time decay of the oscillations of the acoustic modes are markedly affected by the deposition of ML MoS\textsubscript{2}, suggesting a possibly less effective contribution to the thermal transport.

\section{Conclusion}
In this work we used nanoscale EUV transient gratings to comparatively study the nanoscale thermoelastic response of a thin Si membrane and a Si wafer, both covered with ML MoS\textsubscript{2}, and of a blank, thin Si membrane. The results show how a monolayer of a TMD material, in our case MoS\textsubscript{2}, can dramatically modify the response of the Si substrate close to its surface. In particular, ML MoS\textsubscript{2} leads to the almost complete suppression of the surface acoustic wave, which involves elliptical motion localized in proximity of the surface, for comparable excitation fluences. Differently, the longitudinal wave, which does not involve a motion perpendicular to the surface, appears to be less affected, albeit more quickly damped as well. The two Si substrates used, a membrane and a wafer, gave a qualitatively similar behavior. Finally, we observed a slower decay of the non-oscillatory response in the heterostructure, when compared to the blank substrate. This effect is possibly connected to the impact of such acoustic modes on the transport properties of the system at the nanoscale.
This work extends the time-domain investigations of the role of Van der Waals materials on the acoustic response of heterostructures, beyond the case of coherently excited thin layer breathing modes investigated, \textit{e.g.} by time-resolved Brillouin spectroscopy \cite{Vialla2020}{}.

\section{Methods}
\subsection{Samples}
Our study focused on samples based on a 200-nm-thick, 1.6x1.6-mm\textsuperscript{2}-large boron-doped Si membrane, commercially produced by Norcada \cite{Norcada}{}, nominally having $<$ 1 nm rms surface roughness on both front and back sides of the membrane, supported by a 200-\textmu m-thick Si frame.
The three samples are depicted in Fig. \ref{fig1}(a).
One blank Si membrane was used as a reference (sample \#1).
On top of a nominally identical membrane from the same manufacturer, a large-area monolayer of MoS\textsubscript{2} was prepared using the gold tape method \cite{Liu2020}{} (sample \#2) at Columbia University. 
Additionally, we studied a CVD-grown MoS\textsubscript{2} monolayer, deposited on a 0.52-mm-thick Si wafer, commercially available from SixCarbon Technology (sample \#3).

\subsection{Micro-Raman characterization}
The uniformity and thickness of the MoS\textsubscript{2} films were tested with a micro-Raman Renishaw inVia instrument using a continuous-wave 532 nm laser at the I-LAMP laboratories in Brescia (Italy).
Measurements were acquired under an incident power of 5 mW for the samples with a deposited MoS\textsubscript{2} monolayer and 0.5 mW for the blank substrates. At focus, under 100x magnification, the spot size is of the order of a few microns.
All the measurements were performed in air.

\subsection{EUV transient-grating}
The EUV gratings were generated at the EIS-TIMER beamline \cite{Mincigrucci2018,Bencivenga2015,Bencivenga2019}{} using the radiation generated by the double stage FEL source (FEL2) available at the FERMI free-electron laser radiation \cite{Allaria2013}{}. Fig. \ref{fig1}(c) illustrates the EUV TG experimental geometry. Two symmetric pump beams with wavelength $\lambda_{pu}=39.9$ nm and time duration $<$ 100 fs, generated by the first stage of the FEL, are crossed at the sample with an angle $2\theta$=27.6$^{\circ}$ generating a TG with a periodicity $\Pi=2\pi/k_{TG}$=83.6 nm, where  $k_{TG}=4\pi sin(\theta)/\lambda_{pu}=0.075$ nm\textsuperscript{-1} is the modulus of the TG vector. As a probe, we used the output of the second stage of the FEL, which was set at $\lambda_{pr}=13.3$ nm.

The spot sizes, full width at half maximum (FWHM), of the two pump beams were (380x290) \textmu m\textsuperscript{2} and (470x250) \textmu m\textsuperscript{2}. The probe beam at the sample had a spot size of (450x330) \textmu m\textsuperscript{2}. Unlike a typical pump-probe experiment, in a TG measurement the recorded signal originates only from the probe beam fraction which is diffracted by the transient grating. Thus, it only results from the region where the three beams are superimposed.
All the measurements were performed in reflection geometry, in high vacuum and at room temperature. The diffracted signal was recorded using a CCD camera.
The fluence values reported in the text refer to the sum of the incident fluences of the two pumps, whose pulse energies are approximately equal. The FEL intensity at the sample was varied by using a nitrogen gas cell, which efficiently attenuates the radiation at $\lambda_{pu}$ while it marginally affects the transmission of the probe beam. Typically, the acquisition of the EUV-TG signal takes about 1-2 minutes per delay point, leading to 1-2 hours for a full trace and a good signal-to-noise ratio.
An estimate of the uncertainty over the TG signal values at EIS-TIMER beamline was discussed by Bencivenga et al. \cite{Bencivenga2019}{}.
The spatial and temporal overlap between the two pump and the probe beams were checked separately using a reference IR laser and performing FEL-pump/IR-probe transient transmissivity experiments on a YAG reference sample.

The attenuation lengths for the incident FEL pumps in MoS\textsubscript{2} and Si can be derived from the tabulated atomic data found in Henke et al.\cite{Henke1993}{}. We obtain $\approx$40 nm and $\approx$237 nm, respectively, while these values are $\approx$205 nm and $\approx$583 nm at $\lambda_{pr}=13.3$ nm. An extended wavelength dependence and comparison with other common substrates for MoS\textsubscript{2} can be found in the Supplemental Material \cite{supplementary}{}.

\begin{acknowledgement}
A.M. and C.G. acknowledge financial support from MIUR through the PRIN 2015 (Prot. 2015C5SEJJ001) and PRIN 2017 (Prot. 20172H2SC4\_005) programs and from the European Union - Next Generation EU through the MUR-PRIN2022 (Prot. 20228YCYY7) program.

C.G. acknowledges support from Università Cattolica del Sacro Cuore through D.1, D.2.2 and D.3.1 grants. 

Sample fabrication of sample \#2 was supported by the Columbia Nano Initiative and Columbia Materials Science and Engineering Research Center (MRSEC) through NSF grant DMR-2011738.

P.M and S.U. acknowledge funding from the Danish Council for Independent Research, Natural Sciences under the Sapere Aude program (Grant No. DFF- 9064-00057B), VILLUM FONDEN under the Centre of  Excellence for Dirac Materials (grant 11744), and from the Novo Nordisk Foundation (Project Grant NNF22OC0079960).

\end{acknowledgement}

\clearpage
\renewcommand{\thefigure}{S\arabic{figure}}
\captionsetup[figure]{labelsep=period,name={FIG.}, format={plain},justification={raggedright}}
\onecolumn
\begin{center}\Huge Supplementary Information on Impact of a MoS\textsubscript{2} monolayer on the nanoscale thermoelastic response of silicon heterostructures
\end{center}
\vspace{1 cm}

\normalsize

\setcounter{figure}{0} 

\section{Micro-Raman characterization}
\begin{figure}[h]
\includegraphics[width=9 cm]{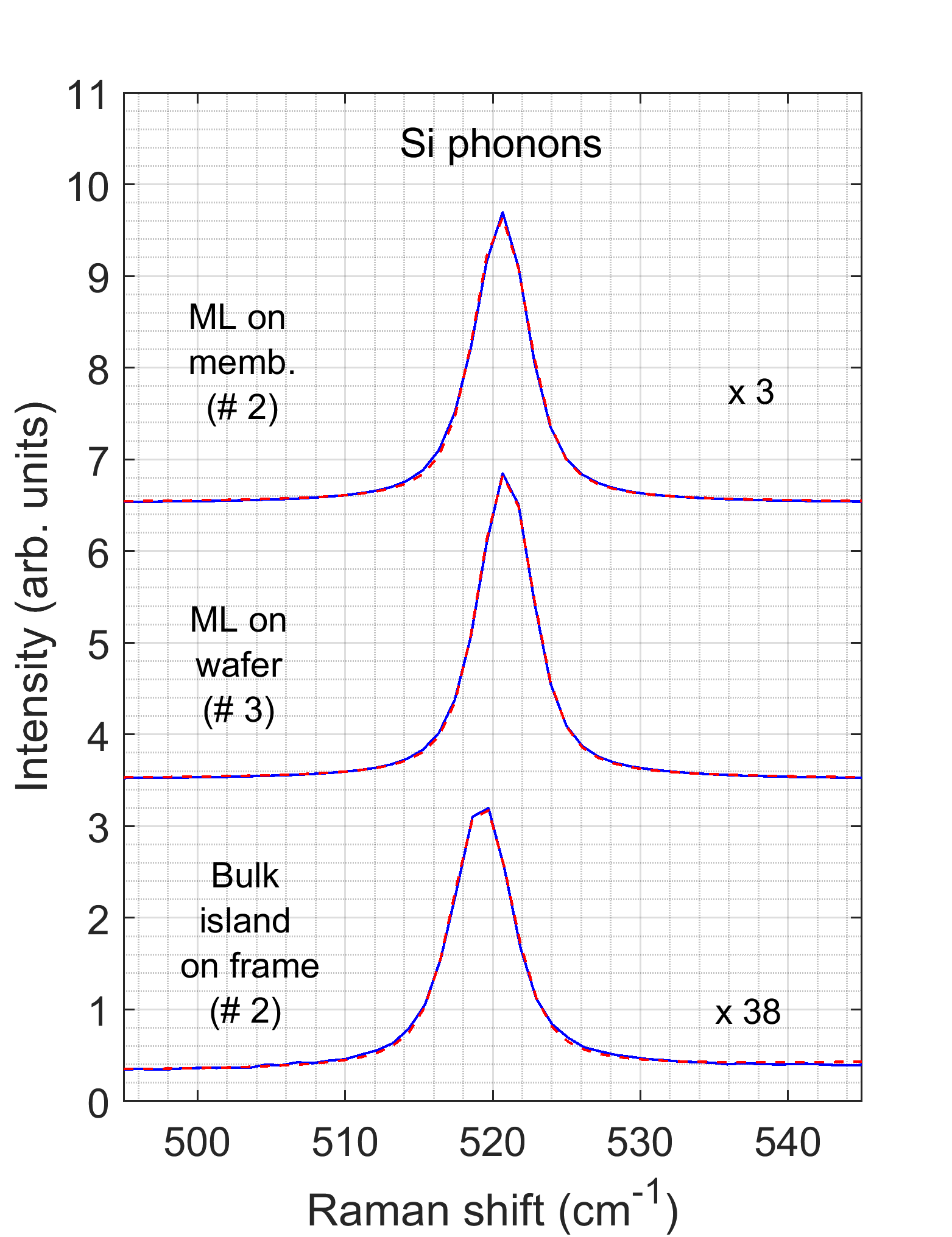}

\caption{Selected micro-Raman spectra showing the Si optical phonon peak from the investigated samples (MoS\textsubscript{2}/Si membrane, MoS\textsubscript{2}/Si wafer and a MoS\textsubscript{2} bulk island). The Raman traces were rescaled and vertically shifted for clarity. The blue lines correspond to the experimental data, while the red-dotted ones to the best fit to a Voigt peak on top of a quadratic background.}

 \label{sfig61}
\end{figure}

In the main text, selected scans of the micro-Raman characterization for the Stokes scattering signal were reported in Fig. 1(b). The spectral window was centered around two peaks corresponding to the $E^1_{2g}$ and $A_{1g}$ near-zone center phonon modes of the MoS\textsubscript{2} film with frequencies at $\approx$385 cm\textsuperscript{-1} and $\approx$404 cm\textsuperscript{-1} respectively. In Fig. \ref{sfig61} we report, measured at the same sample positions, the peaks corresponding to the (degenerate) optical phonon mode of Si at $\approx$520 cm\textsuperscript{-1}.  For sample \#3, however, the Si line was saturating the detector and thus it was acquired in a separate scan with the incident power reduced by a factor 10.
The Raman frequency and narrow linewidth of the Si phonon peak from the substrates confirm a single-crystal structure \cite{Lambertz1998,Yeh2017}{}.

To test the uniformity of the MoS\textsubscript{2} film coverage for samples \#2 and \#3, we performed a 5x5 raster scan, \textit{i.e.}, a series of consecutive Raman spectra taken along a regular grid.

\begin{figure}[h]
\includegraphics[width=\textwidth]{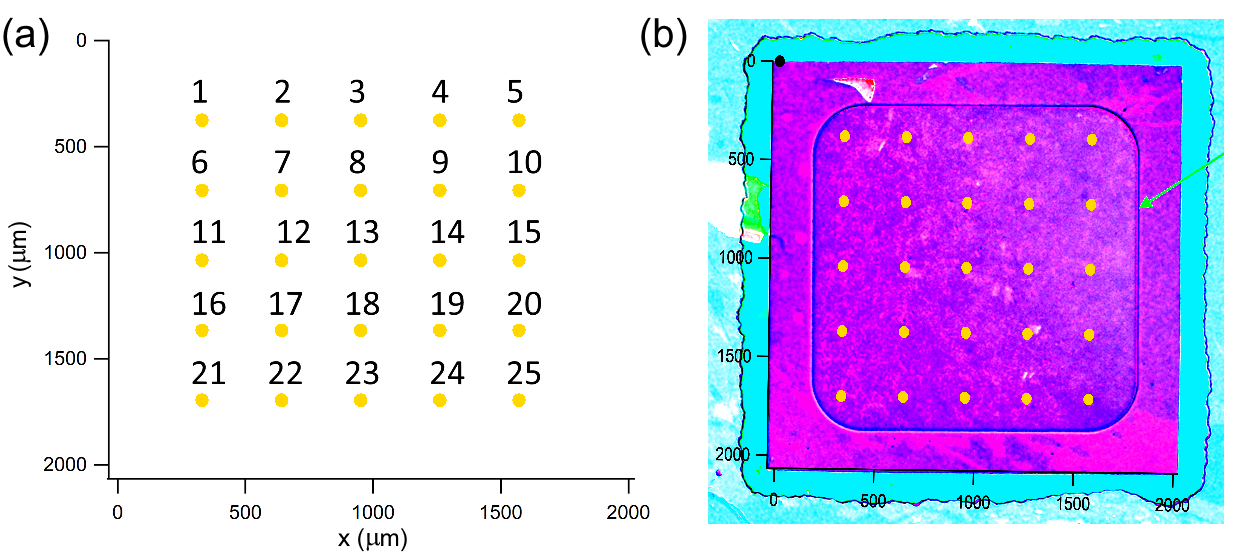}

\caption{(a) Scheme for the raster scan pattern to test the uniformity of the MoS\textsubscript{2} on the silicon substrate. The axes are in micrometers. (b) Optical-microscope picture of sample \#2 after changing the tonality and saturation levels to emphasize the macroscopic morphology of the film. The light blue section corresponds to the frame, while the pink regions to the membrane. The membrane border is indicated by the green arrow.}

 \label{sfig1}
\end{figure}

\begin{figure}[h]
\includegraphics[width=\textwidth]{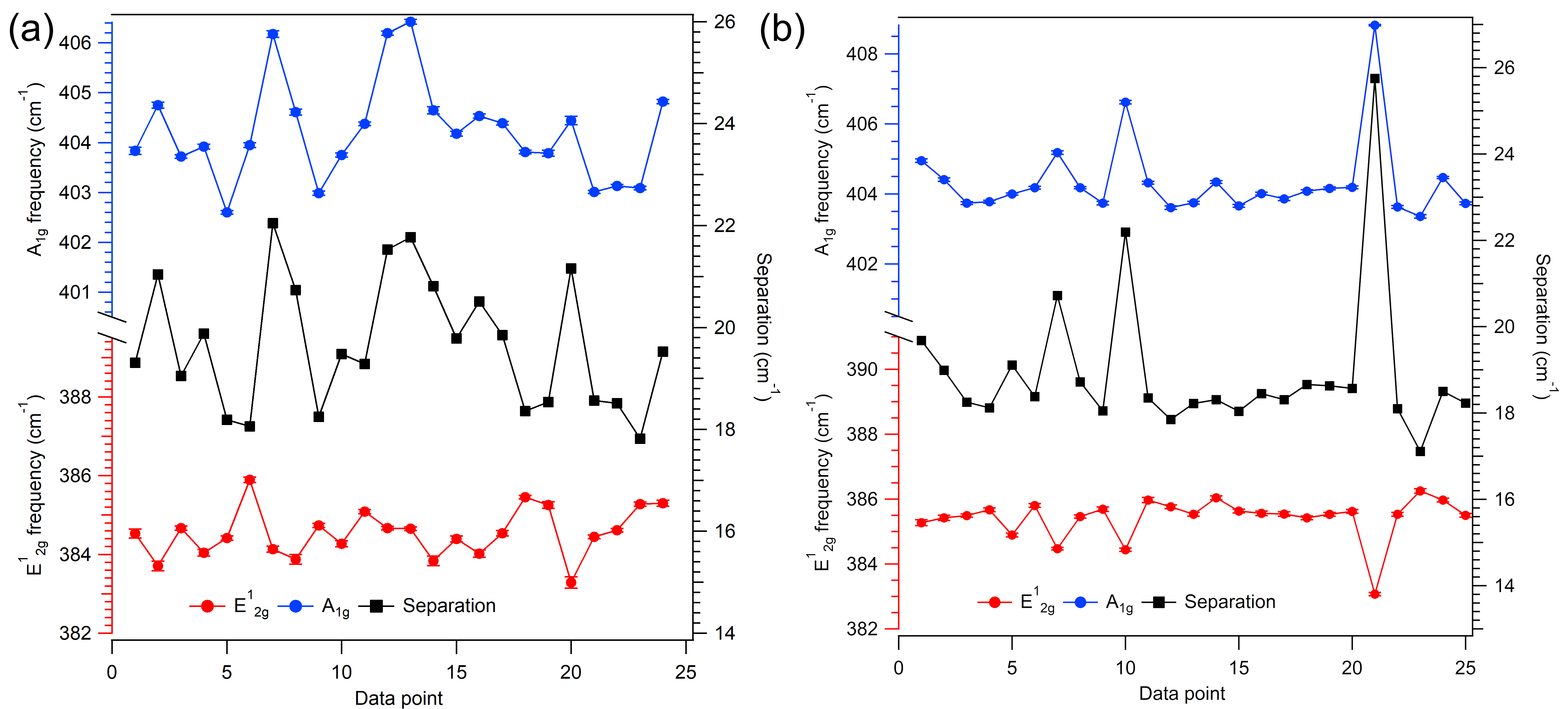}

\caption{The left vertical scale refers to the peak positions for the $E^1_{2g}$ and $A_{1g}$  MoS\textsubscript{2} modes acquired at the positions of the raster scans (Fig. \ref{sfig1}) through micro-Raman measurements, while the right vertical scale refers to the spectral separation between the two peaks; the linewidth of the Raman peaks was about 4 cm$^{-1}$. (a) Sample \#2 (b) Sample \# 3.}

 \label{sfig2}
\end{figure}

In Fig. \ref{sfig1}, we show the 5x5 raster scan scheme employed for the micro-Raman measurements. Panel (a) shows the coordinates of the employed grid (in \textmu{}m), superimposed to sample \#2 in panel (b). An equivalent pattern was used to characterize samples \#1, \#3 and a blank Si wafer.
The spectra were fitted using a series of Voigt peaks, one for each Raman resonance on top of a quadratic background.
The point by point results for the peak frequencies of the two MoS\textsubscript{2} Raman-active modes $E^1_{2g}$ and $A_{1g}$ and their separation are shown in Fig. \ref{sfig2}. In both sample \#2 and \#3, we observe an overall uniform coverage with the values matching those of a monolayer film \cite{Lee2010}{}.

\begin{figure}[h]
\includegraphics[width=9 cm]{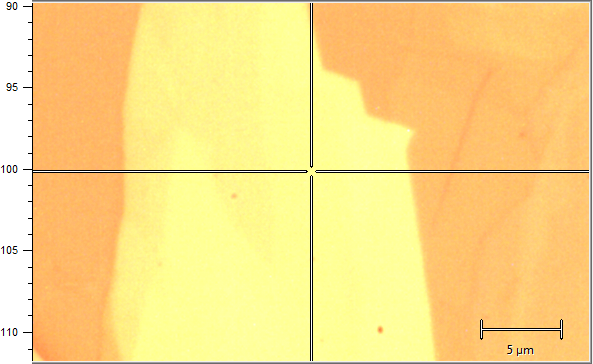}

\caption{Multilayer, showing bulk-like response, MoS\textsubscript{2} region found on the Si frame of sample \#2.}

 \label{sfig12}
\end{figure}

A picture of the multilayer, bulk-like, region found on the Si frame of sample \#2 is reported in Fig. \ref{sfig12}. On such island we acquired the `bulk island' traces from Fig. 1(b) and \ref{sfig61}.

The average frequency ($f$) and FWHM ($w$) of the MoS\textsubscript{2} and Si peaks for the studied samples are summarized in Tables \ref{tab1}, \ref{tab2}, together with the spectral separation of the MoS\textsubscript{2} peaks ($\Delta$).

\begin{table}[h!]
\begin{adjustbox}{width=\columnwidth,center}

\small
\begin{tabular}{ |c|c|c|c|c|c|c|c| } 

\hline
Sample & $f_{E^1_{2g}}$ & $w_{E^1_{2g}}$ & $f_{A_{1g}}$ & $w_{A_{1g}}$ & $\Delta_{A_{1g}-E^1_{2g}}$  & $f_{Si}$ & $w_{Si}$ \\
\hline
MoS\textsubscript{2}/Si m. & $384.6 \pm 0.1$ & $4.02 \pm{} 0.08$ & $404.2 \pm 0.2$ &  $6.0 \pm{} 0.1$ & $19.7 \pm{} 0.3$ & $519.97 \pm{} 0.08$ & $4.61 \pm{} 0.03$ \\ 
\hline
MoS\textsubscript{2}/Si i. & $381.83 \pm{} 0.04$ & $3.83 \pm{} 0.07$ & $407.03 \pm{} 0.02$ &  $4.57 \pm{} 0.04$ & $25.20 \pm{} 0.05$ & $519.29 \pm{} 0.01$ & $4.94 \pm{} 0.02$ \\ 
\hline
MoS\textsubscript{2}/Si w. & $385.4 \pm{} 0.1$ & $4.26 \pm{} 0.04$ & $404.4 \pm{} 0.2$ &  $6.4 \pm{} 0.2$ & $18.9 \pm{} 0.4$ & $520.91 \pm{} 0.04$ & $4.55 \pm{} 0.05$ \\ 
\hline

\end{tabular}
\caption{Micro-Raman characterization results for regions with MoS\textsubscript{2} film. All the numerical results are expressed in cm\textsuperscript{-1}. The abbreviation `m.', `i.' and `w.' stand for `membrane', `island' and `wafer' respectively.}
\label{tab1}

\end{adjustbox}
\end{table}

\begin{table}[h!]
\begin{adjustbox}{width= 6.5 cm,center}

\small
\begin{tabular}{ |c|c|c| } 

\hline
Sample & $f_{Si}$ & $w_{Si}$ \\
\hline

Si membrane & $520.5 \pm{} 0.1$ & $4.32 \pm{} 0.04$ \\ 
\hline
Si wafer &  $520.98 \pm{} 0.06$ & $4.63 \pm{} 0.04$ \\ 
\hline

\end{tabular}
\caption{Micro-Raman characterization results for the filmless substrates. All the numerical results are expressed in cm\textsuperscript{-1}.}
\label{tab2}

\end{adjustbox}
\end{table}
\clearpage

\section{EUV/ Soft x-ray  beam attenuation length}

\begin{figure}[h]
\includegraphics[width=14 cm]{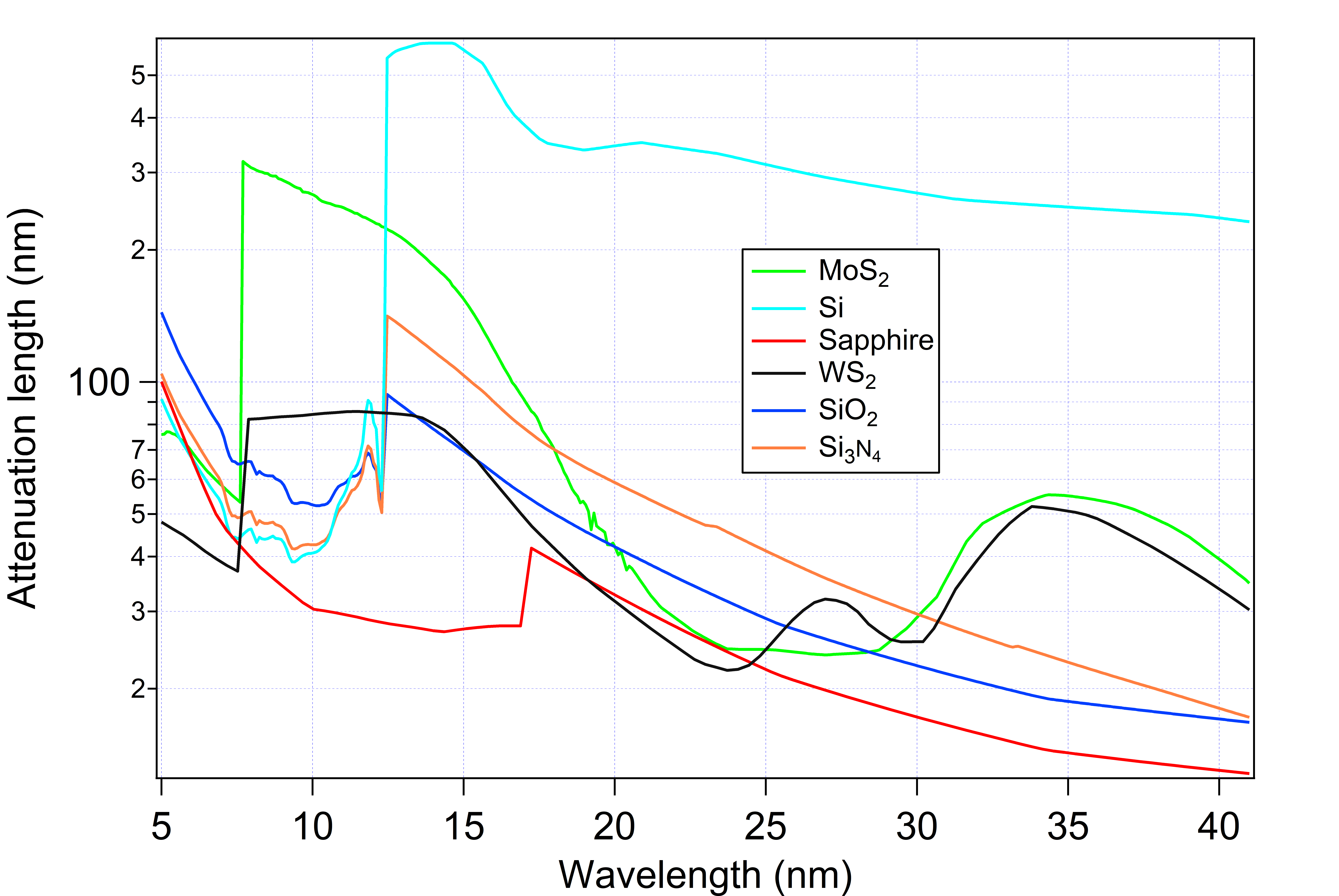}

\caption{EUV / Soft x-ray attenuation lengths for different materials. The curves were calculated using the scattering factor data from \cite{Henke1993}{}.}

 \label{sfig6}
\end{figure}

The EUV / Soft x-ray  wavelength dependence of the attenuation length for different materials is reported in Fig. \ref{sfig6}. As shown by the logarithmic plot, Si is the most transparent one under the wavelengths used in our transient grating FEL experiments (13.3-39.9 nm). At $\lambda_{pu}$=39.9 nm, its attenuation length is about one order of magnitude larger than the others, which indicates that the pump beams travel through the whole Si membrane used for the MoS\textsubscript{2}/Si membrane sample ($\approx$237 nm) without severe energy reduction.
A different case is, e.g., sapphire for which the attenuation length is $\approx$13 nm, therefore meaning that a transient grating generated by 39.9 nm beams will be localized close to the surface and leading to a much higher energy density under the same excitation conditions.

\section{Additional TG results}

\subsection{MoS\textsubscript{2}/Si wafer TG at different points}
\begin{figure}[h!]
\includegraphics[width=9 cm]{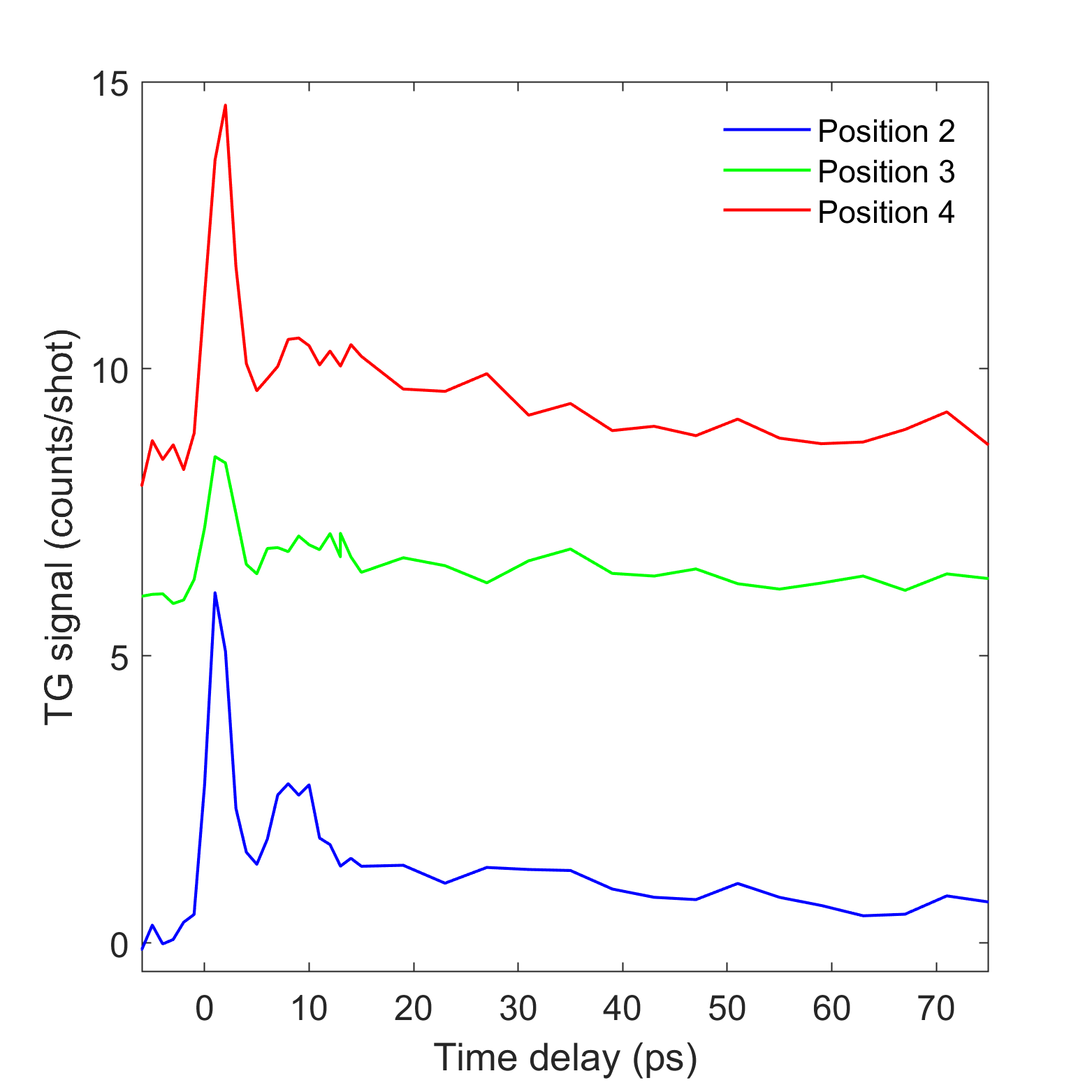}

\caption{TG signal from the MoS\textsubscript{2}/Si wafer sample taken at different positions from the dataset of Fig. 4(a) at $\approx$0.3 mJ/cm\textsuperscript{2} fluence.}

 \label{sfig35}
\end{figure}

We  report in Fig. \ref{sfig35} the TG response collected on the MoS\textsubscript{2}/Si wafer at different positions on the sample. While we observed some variability in the amplitude and decay time of the non-oscillating signal in the recorded window, possibly due to the local morphology of both the substrate and the film, the double peak structure and the general characteristics of the response are maintained.

\subsection{Silicon frame}
\begin{figure}[h]
\includegraphics[width=9 cm]{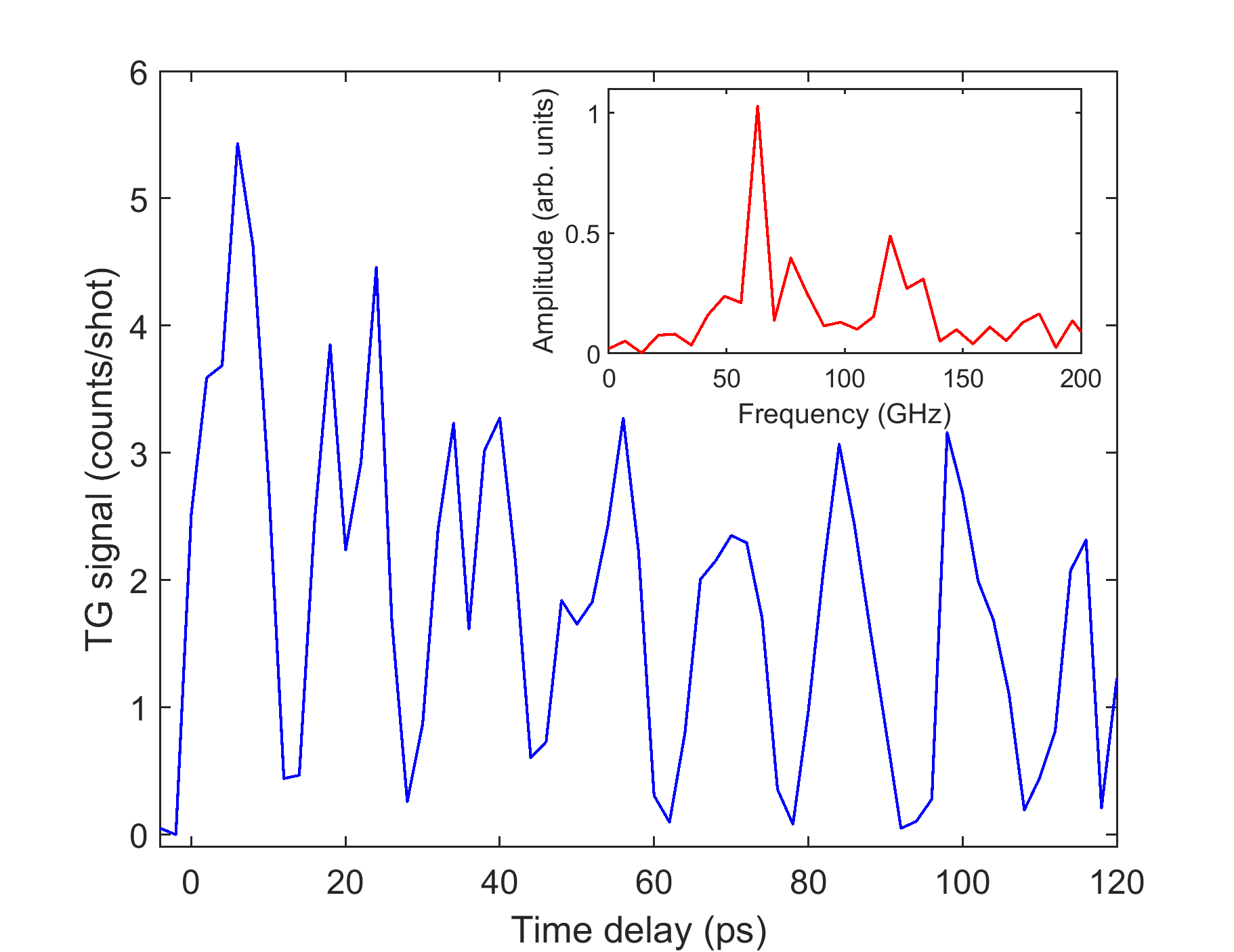}

\caption{Transient grating signal acquired on the Si frame of the blank Si membrane sample using a $83.6$ nm grating period acquired under 0.2 mJ/cm\textsuperscript{2} incident fluence respectively. The inset show the discrete Fourier transform of the TG signal.}

 \label{sfig4}
\end{figure}

As a comparison, we show in Fig. \ref{sfig4} the TG signal obtained from the 200-\textmu m-thick supporting frame of the silicon membrane. While its bulk is still made of silicon, the surface is covered by a thin oxide layer for surface protection purposes.

Compared to the blank membrane response (Fig. 2(a)), the oscillatory response displays a slower decay. 
\subsection{WS\textsubscript{2}/Sapphire}
\begin{figure}[h]
\includegraphics[width=\textwidth]{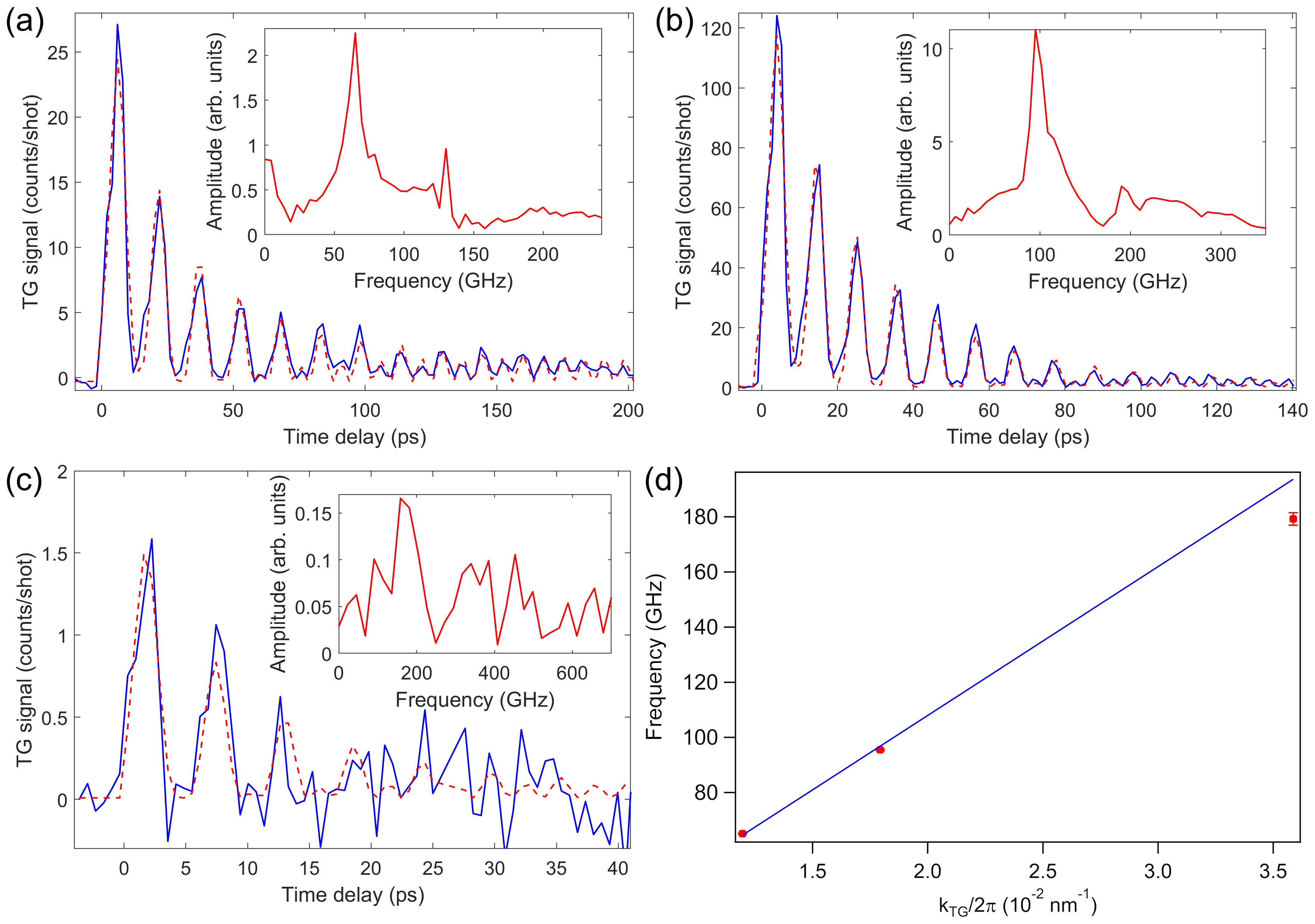}

\caption{TG signal for WS\textsubscript{2}/Sapphire using different transient grating periods. (a) 84 nm under 0.1 mJ/cm\textsuperscript{2} (b) 56 nm under 0.7 mJ/cm\textsuperscript{2} (c) 28 nm under $<$0.1 mJ/cm\textsuperscript{2} (exact estimation not accurately available). (d) Frequency of the Rayleigh mode as a function of the transient grating wavevector $k_{TG}$. The experimental data is represented by blue traces, the fit curves by dotted red lines.}

 \label{sfig45}
\end{figure}

Fig. \ref{sfig45} shows in the EUV TG signal of a different heterostructure, a CVD-grown monolayer WS\textsubscript{2} on a $\approx$0.5-mm c-cut sapphire substrate (commercially available from 2D Semiconductors), collected at different grating periods, obtained by changing $\lambda_{pu}$ from (a) 39.9 nm ($L_{TG}$=83.6 nm) to (b) 26.6 nm ($L_{TG}$=55.8 nm) or (c) 13.3 nm ($L_{TG}$=27.9 nm). Differently from the MoS\textsubscript{2}/Si samples examined in the main text, here we notice that a single oscillatory term in Eq. 1 is enough to describe the TG response. 
In panel (d) we show the obtained frequencies as a function of the transient grating wavevector. Although collected at different fluences, we assume the acoustic wave propagation velocities to be almost independent from it, similarly to the MoS\textsubscript{2}/Si wafer data in Fig. 4(a). Through a linear fit, fixing the intercept to 0, we obtain an estimate propagation speed v=$(5.398 \pm 0.009)$ km/s. This is in good agreement with the Rayleigh mode of sapphire, reported to have a velocity $\approx$5.4 km/s \cite{Pedros2004}{}. Correspondingly, we have for the non-oscillating background decay constant $\tau_{th}$ the following fit results: (a) $(38\pm2)$ ps (b) $(35\pm2)$ ps (c) $(11\pm3)$ ps. The last value is, however, affected by the poorer TG signal signal-to-noise ratio connected to the 13.3 nm pump - 13.3 probe signal, since this configuration is limited by the impossibility to freely vary the relative fluence for pump and probe.
While in the second part of the explored time delay range (especially visible in panels (a) and (b)), we observe a second harmonic modulation, connected to the square in Eq. 2, the main frequency dominates elsewhere with a possible contribution of the longitudinal-like mode the first tens of picoseconds, similarly to the samples studied in the main text. The longitudinal speed in sapphire is expected to be $\approx$11 km/s \cite{Winey2001}{}, corresponding to frequencies of (a) $\approx$132 GHz (b) $\approx$198 GHz (c) $\approx$395 GHz.
The reason why the Rayleigh mode for this sample dominates the response even in the presence of the WS\textsubscript{2} film is likely connected to the much shallower penetration of the pump beams in sapphire (Fig. \ref{sfig6}) compared to Si, leading to a significantly higher energy density near the surface, \textit{i.e.} where the surface acoustic waves such as the Rayleigh mode are localized. Moreover, as we saw in Fig. 4(a) for the  MoS\textsubscript{2} / Si wafer sample, higher excitation fluences could favor the emergence of the Rayleigh mode in the TG signal in WS\textsubscript{2}/Sapphire due to larger initial displacements.

\section{Acoustic waves schemes}

\begin{figure}[h!]
\includegraphics[width=\textwidth]{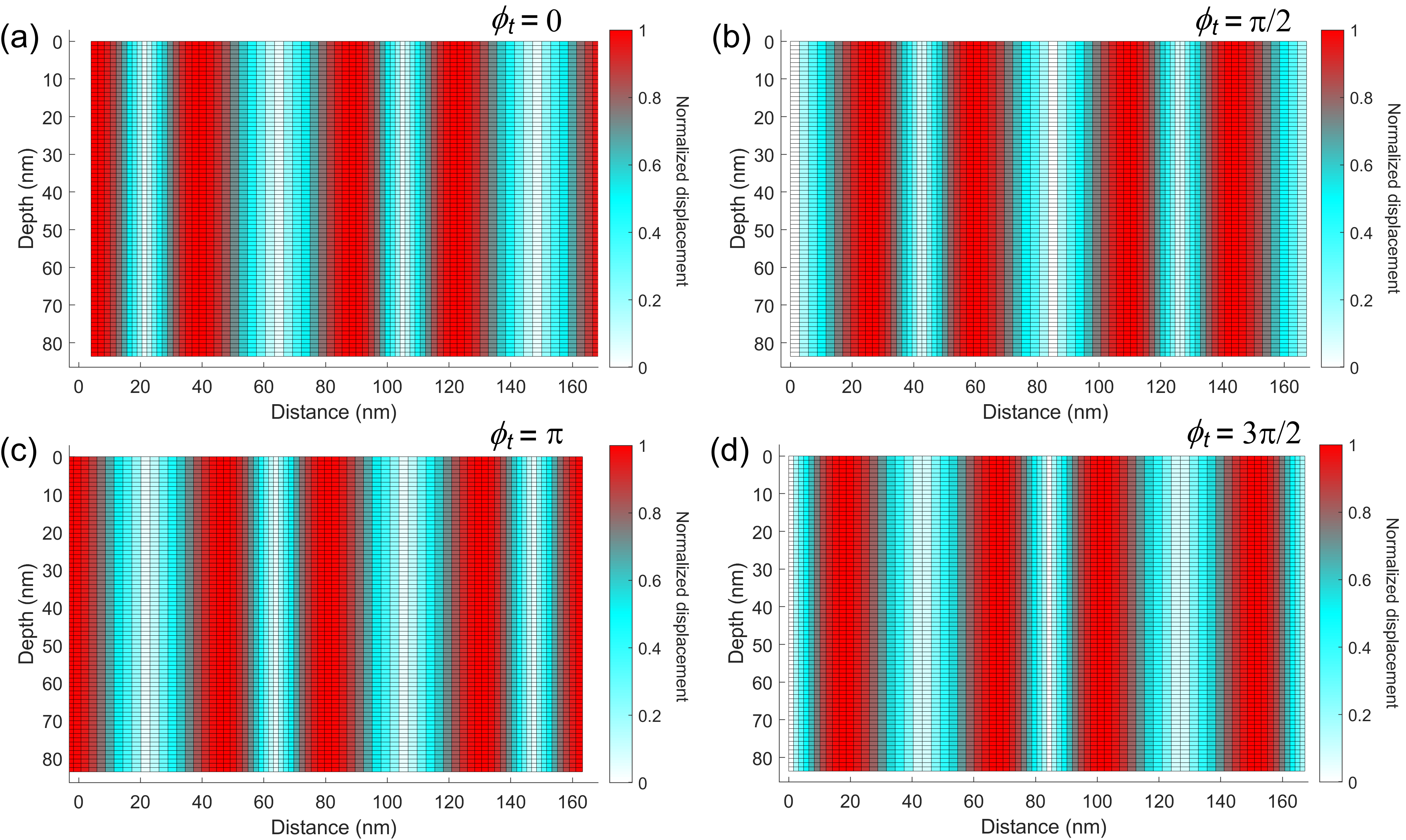}

\caption{Schemes of the displacements connected to a longitudinal wave in Si at four different time instants, expressed as $\phi_t=\omega t$.}

\label{sfig351}
\end{figure}

\begin{figure}[h!]
\includegraphics[width=\textwidth]{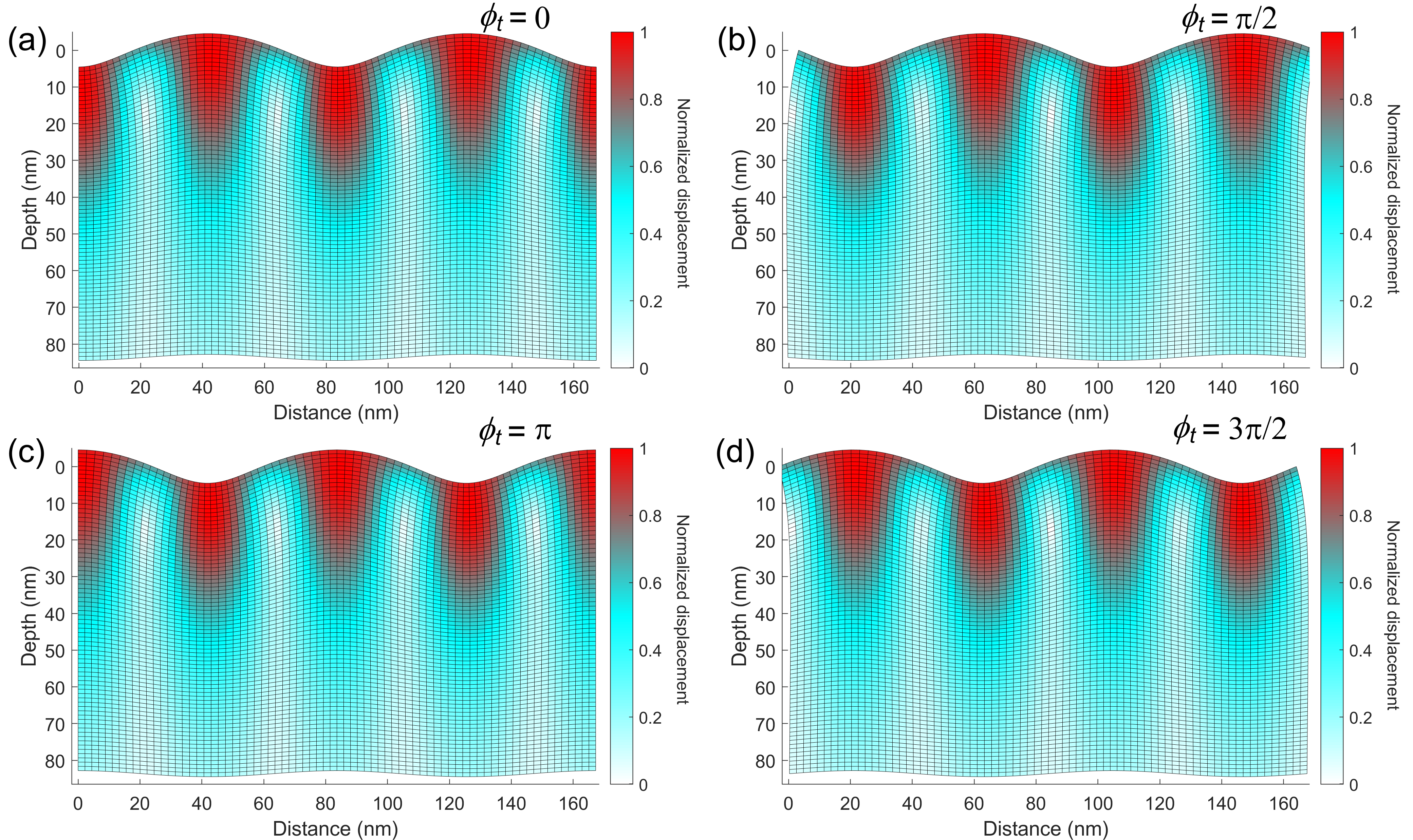}

\caption{Schemes of the displacements connected to a Rayleigh wave in Si at four different time instants, expressed as $\phi_t=\omega t$.}

\label{sfig352}
\end{figure}

In Figs. \ref{sfig351} and \ref{sfig352} we report simulations of the propagation of the longitudinal and Rayleigh mode on a semi-infinite bulk Si substrate at different time delays within an oscillation period.
The plots were drawn using a regular 75 x 75 rectangular point grid. The (relative) displacements were calculated based on the derivation reported in Royer and Valier-Brasier using the method of potentials in the isotopic approximation (Chapter 1 and 3)\cite{Royer2022}{}. The displacements $\vec{u}$ were considered at different positions $x$ parallel to the surface and depths with respect to the surface ($y=0$, positive values for larger depths in Si). For the longitudinal mode propagating parallel to the surface, we used a simple sinusoidal form

\begin{equation}
    \begin{cases}
      u_x(x,y)=U\mathrm{cos}(k_{TG}x-\omega t)\\
      u_y(x,y)=0
    \end{cases}       
\end{equation}
where $U$ is the amplitude of the mode, $k_{TG}$ the wavevector and $\omega$ is the angular frequency. In order to make the shifts visible, we set the amplitude to an arbitrary value. In the plot, the color scale is normalized by the largest magnitude among the calculated displacements.

Regarding the Rayleigh wave, the displacements are
\begin{equation}
    \begin{cases}
      Re[u_x(x,y)]=U_1(y)\mathrm{sin}(k_{TG}x-\omega t)\\
      Re[u_y(x,y)]=U_2(y)\mathrm{cos}(k_{TG}x-\omega t)\\
    \end{cases}       
\end{equation}
where $\omega=k_{TG}v_R$ with $v_R$ Rayleigh velocity and the amplitudes 
\begin{equation}
    \begin{cases}
      U_1(y)=a_L \left[ -k_{TG}e^{-\alpha_Ly}+\sqrt{\alpha_L\alpha_T}e^{-\alpha_T y} \right ] \\
      U_2(y)=a_L\sqrt{\frac{\alpha_L}{\alpha_T}}\left[ -\sqrt{\alpha_L\alpha_T}e^{-\alpha_L y}+k_{TG}e^{-\alpha_Ty} \right] \\
    \end{cases}       
\end{equation}
where $a_L$ is the longitudinal amplitude (connected to the transverse one via $a_T=i\sqrt{\frac{\alpha_L}{\alpha_T}}a_L$), $\alpha_i=\sqrt{k_{TG}^2-k_i^2}$ is a constant with $k_i=\omega/v_i$ connected to the velocity $v_i$ with $i=l,t$ differentiating longitudinal and transverse parameters  respectively. The longitudinal, transverse velocities and Rayleigh values were taken from Farnell and Adler \cite{Farnell1972}{}. To ease visibility, the amplitude $a_L$ was set to an arbitrary value.

\clearpage
\bibliography{biblio_arxiv}

\end{document}